\documentclass{article}
\usepackage{amsmath}
\usepackage{cite}
\usepackage{graphicx}
\usepackage{dcolumn}

\begin{document}

\date{}
\title{The family of Tietz potentials}
\author{Francisco M. Fern\'{a}ndez\thanks{%
fernande@quimica.unlp.edu.ar} \\
INIFTA, DQT, Sucursal 4, C. C. 16, \\
1900 La Plata, Argentina}
\maketitle

\begin{abstract}
We show that several potential-energy functions based on the exponential
function $e^{-\alpha r}$, $\alpha > 0$, are either equivalent or identical
to the Tietz potential. Some of those potentials exhibit redundant
parameters that can be removed without affecting their performance for the
description of the vibration-rotation spectra of diatomic molecules.
\end{abstract}

\section{Introduction}

\label{sec:intro}

Several potentials based on the exponential function $e^{-\alpha r}$, $%
\alpha >0$, have been proposed in the last decades for the treatment of the
vibration-rotation spectra of diatomic molecules. In two papers\cite
{F26,F26b} we have shown that most of those potentials are equivalent to the
Tietz one. To prove this fact we resorted to what we have called
normalization method that consists in rewriting a given potential in terms
of the equilibrium bond length $r_{e}$ and the dissociation energy $D_{e}$.
The choice of a particular, well defined, normal form for all potentials
facilitated their comparison no matter their intricacy. In addition to prove
that all those potentials are equivalent to the Tietz one, we have also
shown that most of them exhibit redundant parameters.

A question arises if those potentials are equivalent only when they are
written in terms of $r_{e}$ and $D_{e}$; that is to say, if they are
required to have a minimum. In this paper we will answer this question by
direct comparison of the potentials.

\section{Normalization method}

\label{sec:Normalization}

For the sake of clarity, we first review the normalization method\cite
{F26,F26b}. Consider a potential $V(r)$, $r>0$, that depends on two or more
adjustable parameters. We require that this potential exhibits a minimum at $%
r=r_{e}$ and a finite value when $r\rightarrow \infty $ (like most commonly
accepted molecular potentials). In order to achieve this goal we force the
given analytical function to satisfy the obvious equations
\begin{equation}
V^{\prime }\left( r_{e}\right) =0,\;\lim\limits_{r\rightarrow \infty
}V(r)-V\left( r_{e}\right) =D_{e}>0,  \label{eq:re,De}
\end{equation}
from which we obtain two model parameters in terms of $r_{e}$, $D_{e}$ and
the remaining potential parameters. If $r_{e}$ is the minimum of the
potential ($V^{\prime \prime }\left( r_{e}\right) >0$) we then rewrite it in
normal form\cite{F26,F26b}
\begin{equation}
V(r)-V\left( r_{e}\right) =D_{e}\left[ 1-f(r)\right] ^{2}.  \label{eq:V_f}
\end{equation}
Since $\lim\limits_{r\rightarrow \infty }\left[ 1-f(r)\right] ^{2}=1$ we
conclude that the two roots $f_{0}(r)$ and $f_{2}(r)$ of the quadratic
equation (\ref{eq:V_f}) behave asymptotically as $\lim\limits_{r\rightarrow
\infty }f_{0}(r)=0$ and $\lim\limits_{r\rightarrow \infty }f_{2}(r)=2$. We
always choose $f_{0}(r)$ in order to provide a consistent representation of $%
V(r)$ in normal form.

As an example, consider the polynomial potential
\begin{equation}
V_{\text{P}}(r)=a+\frac{b}{e^{\alpha r}+d}+\frac{c}{\left( e^{\alpha
r}+d\right) ^{2}},\;\alpha >0,  \label{eq:V_P}
\end{equation}
where $d\geq -1$ in order to avoid an unphysical singularity in $0<r<\infty $%
. Clearly, $a$ is an irrelevant energy shift that does not appear in
equations (\ref{eq:re,De}); however, it will be useful for comparing $V_{%
\text{P}}(r)$ with other potentials. If we solve equations (\ref{eq:re,De})
for $b$ and $c$ we obtain
\begin{equation}
b=-2D_{e}\left( e^{\alpha r_{e}}+d\right) ,\;c=D_{e}\left( e^{\alpha
r_{e}}+d\right) ^{2},  \label{eq:b,c_P}
\end{equation}
and the normal form
\begin{equation}
V_{\text{P}}(r)-V_{\text{P}}\left( r_{e}\right) =V_{N}(r)=D_{e}\left( 1-%
\frac{e^{\alpha r_{e}}+d}{e^{\alpha r}+d}\right) ^{2},  \label{eq:V_P_normal}
\end{equation}
of the potential (\ref{eq:V_P}).

If we define $k=V^{\prime \prime }\left( r_{e}\right) $ and $k_{3}=V^{\prime
\prime \prime }\left( r_{e}\right) $ then the remaining model parameters can
be determined from the following equations
\begin{eqnarray}
d &=&e^{\alpha r_{e}}\left( \alpha \sqrt{\frac{2D_{e}}{k}}-1\right) ,
\nonumber \\
\alpha &=&\sqrt{\frac{2k}{D_{e}}}+\frac{k_{3}}{3k}.  \label{eq:d,alpha}
\end{eqnarray}
The polynomial potential contains one irrelevant energy shift $a$ and four
adjustable parameters $b$, $c$, $d$ and $\alpha $ that can be expressed in
terms of $r_{e}$, $D_{e}$, $k$ and $k_{3}$ that can be obtained from
spectroscopic data\cite{H50}.

We say that two potentials $V_{1}(r)$ ad $V_{2}(r)$ are equivalent if $%
V_{1}^{\prime }(r)-V_{2}^{\prime }(r)=0$ for all $r$. The reason is that the
Hamiltonian operators $H=T+V_{i}(r)$, $i=1,2$, where $T$ is the
kinetic-energy operator, have the same eigenfunctions and their eigenvalues
are displaced by the irrelevant energy shift $\Delta =V_{1}(r)-V_{2}(r)$.
Therefore, $V_{1}(r)$ and $V_{2}(r)$ describe the same physical model. If $%
V_{1}(r)$ is equivalent to $V_{2}(r)$ and $V_{2}(r)$ is equivalent to $%
V_{3}(r)$ then $V_{1}(r)$ is equivalent to $V_{3}(r)$ as follow from $%
V_{1}^{\prime }(r)-V_{3}^{\prime }(r)=V_{1}^{\prime }(r)-V_{2}^{\prime
}(r)+V_{2}^{\prime }(r)-V_{3}^{\prime }(r)=0$. A particular case of
equivalent potentials is that of identical potentials for which $\Delta =0$.

According to the definition above $V_{\text{P}}(r)$ is equivalent to $%
V_{N}(r)$ when the potential parameters are given by equation (\ref{eq:b,c_P}%
). In what follows $V_{\text{P}}(r,a,b,c,d,\alpha )$ will denote the
polynomial potential (\ref{eq:V_P}). We keep the irrelevant energy shift $a$
in order to facilitate comparison with other potentials based on exponential
functions because we will look for exact identity. Given a potential $V_{%
\text{G}}(r)$ we will rewrite it as $V_{\text{G}}(r)=V_{\text{P}%
}(r,a,b,c,d,\alpha )$ indicating the corresponding values of $a$, $b$, $c$, $%
d$, and $\alpha $.

\section{Tietz potential}

\label{sec:Tietz}

One of the most relevant potentials for present analysis was proposed by
Tietz\cite{T63} several years ago
\begin{equation}
V_{\text{T}}(r)=D_{e}+D_{e}\frac{\left( a+b\right) e^{-2\beta r}-be^{-\beta
r}}{\left( 1+ce^{-\beta r}\right) ^{2}},\;\beta >0,  \label{eq:V_T}
\end{equation}
where $c\geq -1$ in order to avoid an unphysical singularity in $0<r<\infty $%
.

We can easily prove that the Tietz potential is identical to the polynomial
one:
\begin{eqnarray}
V_{\text{T}}(r) &=&V_{\text{P}}(r,a^{\prime },b^{\prime },c^{\prime
},d^{\prime },\alpha ),\;a^{\prime }=D_{e},\;b^{\prime }=-bD_{e},\;c^{\prime
}=aD_{e}+bD_{e}(c+1),\;  \nonumber \\
d^{\prime } &=&c,\;\alpha =\beta .
\end{eqnarray}

The simple potential
\begin{equation}
V_{S}(r)=\frac{A+Be^{-\alpha r}}{\left( 1-Ce^{-\alpha r}\right) ^{2}}+D,
\label{eq:V_S}
\end{equation}
is also identical to the polynomial one as shown by
\begin{eqnarray}
V_{\text{S}}(r) &=&V_{\text{P}}(r,a,b,c,d,\alpha
),\;a=A+D,\;b=2AC+B,\;c=C(AC+B),\;  \nonumber \\
d &=&-C.
\end{eqnarray}

The potentials $V_{\text{P}}(r)$ and $V_{\text{S}}(r)$ are the simplest
members of the Tietz family.

\section{Other potentials based on exponential functions}

\label{sec:Potentials}

The Hua potential\cite{H90}
\begin{equation}
U_{\text{H}}(r)=D_{e}\left[ \frac{1-e^{-b\left( r-r_{e}\right) }}{%
1-ce^{-b\left( r-r_{e}\right) }}\right] ^{2},\;b>0,  \label{eq:V_H}
\end{equation}
is given in terms of four parameters and it also belongs to the Tietz family
\begin{eqnarray}
U_{\text{H}}(r) &=&V_{\text{P}}(r,a^{\prime },b^{\prime },c^{\prime
},d^{\prime },b),\;a^{\prime }=D_{e},\;b^{\prime }=2D_{e}e^{br_{e}}\left(
c-1\right) ,\;  \nonumber \\
c^{\prime } &=&D_{e}e^{2br_{e}}\left( c-1\right) ^{2},\;d^{\prime
}=-ce^{br_{e}}.
\end{eqnarray}

\c{S}im\c{s}ek and \"{O}z\c{c}elik\cite{SO94} proposed the four-parameter
exponential type potential (FPEP)
\begin{equation}
V_{\text{FPEP}}(r)=\frac{Ae^{\lambda r}}{C+e^{\lambda r}}-\frac{Be^{\lambda
r}}{\left( C+e^{\lambda r}\right) ^{2}},\;A,B>0,  \label{eq:V_FPEP}
\end{equation}
that is just another member of the Tietz family
\begin{equation}
V_{\text{FPEP}}(r)=V_{\text{P}}(r,a^{\prime },b^{\prime },c^{\prime
},d^{\prime },\lambda ),\;a^{\prime }=A,\;b^{\prime }=-AC-B,\;c^{\prime
}=BC,\;d^{\prime }=C.
\end{equation}

Jia et al.\cite{JIAETAL01} introduced a deformed five-parameter
exponential-type potential (DFPEP)
\begin{equation}
V_{\text{DFPEP}}(r)=P_{1}+\frac{P_{2}}{e^{2\alpha r}+q}+\frac{P_{3}}{\left(
e^{2\alpha r}+q\right) ^{2}},\;\alpha >0,  \label{eq:V_DFPEP_1}
\end{equation}
that is identical to the polynomial potential

\begin{equation}
V_{\text{DFPEP}}(r)=V_{\text{P}}(r,P_{1},P_{2},P_{3},q,2\alpha ),
\end{equation}
and consequently is a member of the Tietz family.

Boonserm and Visser\cite{BV11} showed that several potentials are equivalent
to the M\"{o}bius square potential
\begin{equation}
V(r)=V_{0}\left( \frac{A+Be^{-\alpha r}}{C+De^{-\alpha r}}\right) ^{2}.
\label{eq:V_Mob}
\end{equation}
It can be rewritten in terms of the polynomial potential as
\begin{eqnarray}
V(r) &=&V_{\text{P}}(r,a,b,c,d,\alpha ),\;a=V_{0}\frac{A^{2}}{C^{2}}%
,\;b=V_{0}\frac{2A\left( BC-AD\right) }{C^{3}},\;  \nonumber \\
c &=&V_{0}\frac{\left( AD-BC\right) ^{2}}{C^{4}},\;d=\frac{D}{C},
\end{eqnarray}
and therefore belongs to the Tietz family.

Falaye et al.\cite{FIH15} proposed the shifted Tietz-Wei (sTW) oscillator
\begin{equation}
V_{\text{sTW}}(r)=V_{e}\frac{2(c_{h}-1)e^{-b_{h}\left( r-r_{e}\right)
}-\left( c_{h}^{2}-1\right) e^{-2b_{h}\left( r-r_{e}\right) }}{\left[
1-c_{h}e^{-b_{h}\left( r-r_{e}\right) }\right] ^{2}},\;b_{h}>0,
\label{eq:V_sTW_1}
\end{equation}
given in terms of four adjustable parameters. It can be rewritten as
\begin{eqnarray}
V_{\text{sTW}}(r) &=&V_{\text{P}}(r,0,b^{\prime },c^{\prime },d^{\prime
},b_{h}),\;b^{\prime }=2V_{e}e^{b_{h}r_{e}}\left( c_{h}-1\right) ,\;
\nonumber \\
c^{\prime } &=&V_{e}e^{2b_{h}r_{e}}\left( c_{h}-1\right) ^{2},\;d^{\prime
}=-c_{h}e^{b_{h}r_{e}},
\end{eqnarray}
which proves that it belongs to the Tietz family.

Onate et al.\cite{OOOI20} proposed the q-deformed six-parameter
exponential-type potential model
\begin{equation}
V_{\text{qDSPEP}}(r)=Q_{1}-\frac{Q_{2}}{1-qe^{-2\alpha r}}+\frac{%
Q_{3}e^{-2\alpha r}}{1-qe^{-2\alpha r}}-\frac{Q_{4}e^{-4\alpha r}}{\left(
1-qe^{-2\alpha r}\right) ^{2}},\;\alpha >0,  \label{eq:V_six-param}
\end{equation}
that is identical to the polynomial potential as shown by
\begin{eqnarray}
V_{\text{qDSPEP}}(r) &=&V_{\text{P}}(r,a,b,c,d,2\alpha
),\;a=Q_{1}-Q_{2},\;b=Q_{3}-qQ_{2},\;c=-Q_{4},\;  \nonumber \\
d &=&-q,
\end{eqnarray}
and also belongs to the Tietz family. Note that if we choose $Q_{2}=0$ or $%
Q_{3}=0$ the performance of the model is not affected. Any one of those
parameters is redundant.

Okorie et al.\cite{OIC20} proposed the deformed exponential-type potential
(DEP)
\begin{equation}
U_{\text{DEP}}(r)=P_{1}+\frac{P_{2}e^{-2\alpha \left( r-r_{0}\right) }}{%
1-qe^{-2\alpha \left( r-r_{0}\right) }}+\frac{P_{3}e^{-4\alpha \left(
r-r_{0}\right) }}{\left[ 1-qe^{-2\alpha \left( r-r_{0}\right) }\right] ^{2}},
\label{eq:U_DEP}
\end{equation}
that is also identical to the polynomial potential
\begin{eqnarray}
U_{\text{DEP}}(r) &=&V_{\text{P}}(r,a,b,c,d,2\alpha
),\;a=P_{1},\;b=P_{2}e^{2\alpha r_{0}},\;c=P_{3}e^{4\alpha r_{0}},\;
\nonumber \\
d &=&-qe^{2\alpha r_{0}},
\end{eqnarray}
and belongs to the Tietz family. In this case we can choose $r_{0}=0$
without affecting the performance of the model. This parameter is redundant.

In their research Onyenegecha et al.\cite{OOEIH20} and Yanar et al.\cite
{YSSA24} resorted to the potential

\begin{equation}
V_{\text{MH}}(r)=-V_{0}\left( \frac{A+Be^{-2\alpha r}}{1-e^{-2\alpha r}}%
\right) ^{2}+\frac{V_{1}e^{-2\alpha r}}{1-e^{-2\alpha r}},
\label{eq:V_Mob+Hult}
\end{equation}
that is also identical to the polynomial potential
\begin{eqnarray}
V_{\text{MH}}(r) &=&V_{\text{P}}(r,a,b,c,-1,2\alpha
),\;a=-V_{0}A^{2},\;b=V_{1}-2V_{0}A(A+B),\;  \nonumber \\
c &=&-V_{0}\left( A+B\right) ^{2},
\end{eqnarray}
and belongs to the Tietz family. $V_{\text{MH}}(r)$ is a three-parameter
model because $d=-1$. In this case we can choose, for example, $V_{1}=0$ and
$A=1$ without affecting the model performance. In other words, two
parameters are redundant.

Yanar et al.\cite{YSSA24} mentioned the apparently more general
six-parameter potential-energy function
\begin{equation}
V_{\text{MH+}}(r)=-U_{0}\left( \frac{A+Be^{-2\alpha r}}{1-e^{-2\alpha r}}%
\right) ^{2}-\frac{U_{1}e^{-2\alpha r}}{1-e^{-2\alpha r}}+\frac{%
U_{2}e^{-2\alpha r}}{\left( 1-e^{-2\alpha r}\right) ^{2}},
\label{eq:V_Mob+Hult_gen}
\end{equation}
that had been used earlier by Njoku et al.\cite{NJOETAL23} and Onyenegecha
et al.\cite{ONYE22}. This potential can also be reduced to the polynomial
potential
\begin{eqnarray}
V_{\text{MH+}}(r) &=&V_{\text{P}}(r,a,b,c,-1,2\alpha
),\;a=-U_{0}A^{2},\;b=U_{2}-U_{1}-2U_{0}A(A+B),\;  \nonumber \\
c &=&U_{2}-U_{0}\left( A+B\right) ^{2},
\end{eqnarray}
and belongs to the Tietz family. It is also a three-parameter model because $%
d=-1$. In this case we can choose $A=1$, $U_{1}=0$ and $U_{2}=0$ without
affecting the model performance. There are three redundant parameters.

Yanar et al.\cite{YAS16} and Yanar\cite{Y22} made extensive use of the
general molecular potential
\begin{equation}
V_{\text{GMP}}(r)=\frac{A-Be^{-\alpha \left( r-r_{e}\right) }+\tilde{q}%
\left[ C-De^{-\alpha \left( r-r_{e}\right) }\right] ^{2}}{\left[
1-qe^{-\alpha \left( r-r_{e}\right) }\right] ^{2}},  \label{eq:GMP}
\end{equation}
that is also identical to the polynomial potential
\begin{eqnarray}
V_{\text{GMP}}(r) &=&V(r,a,b,c,d,\alpha ),\;a=A+\tilde{q}C^{2},\;  \nonumber
\\
b &=&e^{\alpha r_{e}}\left[ 2Aq-B+2C\tilde{q}(Cq-D\right] ,\;  \nonumber \\
c &=&e^{2\alpha r_{e}}\left[ Aq^{2}-Bq+\tilde{q}\left(
D^{2}-2CDq+C^{2}q^{2}\right) \right] ,\;  \nonumber \\
d &=&-qe^{\alpha r_{e}},
\end{eqnarray}
and belongs to the Tietz family. In this case we can choose $\tilde{q}=0$
(or $C=D=0$) and $r_{e}=0$ without affecting the model performance. Note
that when $\tilde{q}=0$ and $r_{e}=0$ we obtain the simple potential (\ref
{eq:V_S}). The parameters $\tilde{q}$, $C$, $D$ and $r_{e}$ are redundant.
Note that $\left. \frac{dV_{\text{GMP}}}{dr}\right| _{r=r_{e}}\neq 0$ and
the equilibrium bond length appears when we use equations (\ref{eq:re,De}).

\section{Conclusions}

\label{sec:conclusions}

The results given above show that many potentials built from the exponential
function $e^{-\alpha r}$ are identical (or at least equivalent) to the Tietz
potential. When used for the study of the vibration-rotation spectra of
diatomic molecules, all of them reduce to an identical expression given in
terms of $r_{e}$ and $D_{e}$ like equation (\ref{eq:V_P_normal}) that can be
complemented by equations (\ref{eq:d,alpha}). None of those potentials
provide any advantage with respect to such physical application.

Some of the potentials discussed in the preceding section exhibit redundant
parameters that can be removed without affecting the performance of the
model. This is the case of the potentials proposed by Onate et al.\cite
{OOOI20}, Okorie et al.\cite{OIC20}, Onyenegecha et al.\cite{OOEIH20}, Yanar
et al.\cite{YSSA24}, Njoku et al.\cite{NJOETAL23}, Onyenegecha et al.\cite
{ONYE22}, Yanar et al.\cite{YAS16} and Yanar\cite{Y22}.

\end{document}